\documentclass[graphics,floatfix, footinbib,tightenlines,nobibnotes, aps, prb, twocolumn]{revtex4-1}
\usepackage{amsmath,amssymb,wasysym}
\usepackage{graphicx}% Include figure files
\usepackage{dcolumn}% Align table columns on decimal point
\usepackage{bm}% bold math
\usepackage{braket}
\usepackage{subcaption}
\usepackage{verbatim}
\usepackage{float}
\usepackage{bbold}
\usepackage{color}
\usepackage{xcolor}
\usepackage{relsize}
\usepackage{amsthm}
\usepackage{enumerate}
\usepackage{soul,xcolor}
\usepackage[T1]{fontenc}
\usepackage[colorlinks=true ,urlcolor=blue,urlbordercolor={0 1 1}]{hyperref}

\newcommand{\beq}{\begin{eqnarray}}
\newcommand{\eeq}{\end{eqnarray}}

\newcommand{\bsp}{\begin{split}}
\newcommand{\esp}{\end{split}}

\newcommand{\be}{\begin{equation}}
\newcommand{\ee}{\end{equation}}

\begin{document}

\setstcolor{red}

\title{Type II $t-J$ model in superconducting nickelate Nd$_{1-x}$Sr$_x$NiO$_2$}
\author{Ya-Hui Zhang and Ashvin Vishwanath}
\affiliation{Department of Physics, Harvard University, Cambridge, MA, USA
}

\date{\today}% It is always \today, today,

\begin{abstract}
The recent observation of superconductivity at relatively high temperatures in  hole doped NdNiO$_2$ has generated considerable interest, particularly due to its similarity with the infinite layer cuprates. Building on the observation that the Ni$^{2+}$ ions resulting from hole doping are commonly found in the spin-triplet state, we introduce and study a variant of the $t-J$ model in which the holes carry S=1. We name this new model the Type II $t-J$ model. We find two distinct  mechanisms for $d$ wave superconductivity. In both scenarios the pairing is driven by the spin coupling $J$.
However, coherence is gained in distinct ways in these two scenarios. In the first case, the spin-one holes
condense  leading to a  $d$ wave superconductor along with spin-symmetry breaking. Different orders including spin-nematic orders are possible.  This scenario is captured by a spin one slave boson theory. In the second scenario, a coherent and symmetric $d$ wave superconductor is achieved from "Kondo resonance": spin one holes contribute two electrons to form large Fermi surface together with the spin 1/2 singly occupied sites. The large Fermi surface then undergoes $d$ wave pairing because of spin coupling $J$, similar to heavy fermion superconductor.  We propose a three-fermion parton theory to treat these two different scenarios in one unified framework and calculate its doping phase diagram within a self consistent mean field approximation.  Our study shows that a combination of "cuprate physics" and "heavy fermion physics" may emerge in the type II $t-J$ model.
\end{abstract}

\pacs{Valid PACS appear here}% PACS, the Physics and Astronomy
\maketitle{}

\textit{\bf{Introduction}}
Recently, a  tour de force materials synthesis effort created a thin-film of the hole doped infinite-layer nickelate NdNiO$_2$\cite{li2019superconductivity}.  In this material, the uncommon  Ni$^{1+}$ in the $3d^9$ configuration is realized, similar to  Cu$^{2+}$ in the high T$_c$ cuprate materials. Remarkably, a relatively high superconducting transition temperature  $T_c \approx 9-15$ K was reported \cite{li2019superconductivity}. Besides, according to LDA+U calculations \cite{Anisimov1999, Chaloupka2008,Pickett2004,Norman2019,Kuroki2019,ZXShen2019,2019arXiv190903942N,gao2019electronic,ryee2019induced,zhang2019effective}, the band at Fermi level is dominated by the $d_{x^2-y^2}$ orbital of Ni, which suggests that the main physics may also be governed by a one-band Hubbard model as in the cuprates.  However, in this paper we argue that the physics of the hole doped nickelate is essentially different from that of the cuprates.  In cuprates,  the near degeneracy of oxygen $2p$ and copper $d_{x^2-y^2}$ orbitals leads to the well known fact that the doped hole enters the oxygen $2p$ orbital and forms the Zhang-Rice singlet \cite{zhang1988effective}.  In contrast, the oxygen $2p$ orbital is  far away from the Fermi level in the nickelate, due to the lower oxidation state of the Ni$^{1+}$ ion compared to Cu$^{2+}$. Therefore the doped hole enters the $3d$ orbital and creates a  Ni$^{2+}$ state with $3d^8$ configuration.  The remaining question is whether the hole is in the low (S=0) or high (S=1) configuration. The Ni$^{2+}$ ion is often found in the high spin $S=1$ state\cite{zaanen1985band, Sawatzky2019}, thanks to Hund's first rule . For example, the spin one Haldane chain is realized in the Ni$^2+$,  $d^8$ configuration\cite{kojima1995musr}. In this case, we expect the  hole doped NdNiO$_2$ to likely be described by a novel $t-J$ model with spin-one holes. However, in the absence of a direct experimental measurement of the spin state of the doped hole, one  cannot rule out the possibility that Ni$^{2+}$ is in the low spin state because of a larger energy splitting of  the two $e_g$ orbitals. Indeed such a low spin configuration was proposed in a different but related compound\cite{zhang2017large} based on certain spectroscopic measurements although more data may be needed to confirm the conclusion.   In this case one must revert to a cuprate-like $t-J$ model\cite{Raghu2019, Sawatzky2019,hirsch2019hole,singh2019road,fu-chun_zhang}, at least as far as doped holes are concerned.  In this paper we study the unconventional $t-J$ model with spin-one holes, which we dub the Type II $t-J$ model. This novel model is of theoretical interest even as its relevance to the hole doped NdNiO$_2$ awaits  experimental confirmation.  Besides, we hope our theoretical analysis will motivate more experimental searches for realizing this Type II $t-J$ model. Doping electrons into the $3d^7$ configuration is also promising in this regard.

{\bf Type II t-J model:} Let us sketch the form of the type II $t-J$ model, more details on the derivation from the  microscopic Hubbard Hamiltonian can be found in Appendix~\ref{append_derive_t_J}. For convenience, we use the hole picture in this paper and define the vacuum as the $3d^{10}$ state for each site, and describe the particle-hole transformed version of the original problem. In this picture, the undoped parent compound has a single hole on each site, which we will call a {\em singlon}, while the state with two holes obtained on doping, will be called a {\em doublon}. The doped hole  enters the $d_{z^2}$ orbital and creates a doubly occupied site with two holes sitting on the two $e_g$ orbitals. Because of the inter-orbital Hund's coupling, the two holes form a spin  triplet (S=1) state. We assume that the singlon is always on the $d_{x^2-y^2}$ orbital because of the splitting between the two $e_g$ orbitals.   Then the Hilbert space at each site consists of two singlon states and three doublon states.  We label the  spin $1/2$ singlon with $\sigma=\uparrow, \downarrow$ and label the triplet doublon with $a=x,y,z$.  We also define a density operator at each site: $n_i= \sum_a \ket{a}\bra{a}$. Thus $n_i$ measures the number of doublons and is equal to the number of doped holes.  At the doping level $x$, the density of singlon and doublon is $1-x$ and $x$ respectively and we have $\langle n_i \rangle=x$.

 The physical spin 1/2 and spin one  operators are ${S^s_a} = \frac{{\sigma^a}_{\sigma\sigma'}}{2}\ket{\sigma}\bra{\sigma'}$ and  ${S^d_a} = -i\epsilon_{abc}\ket{b}\bra{c}$ respectively. {The hole operator for the $d_{x^2-y^2}$ orbital is zero (i.e. has vanishing matrix elements) in the restricted Hilbert space and the only hole operator we have is the one corresponding to the $d_{z^2}$ orbital. Microscopically this hole operator $c^\dagger_\sigma$ creates a hole on $d_{z^2}$ orbital.}   In the restricted Hilbert space it acts as
\begin{align}
c_{i;\uparrow}&= \prod_{j<i}(-1)^{n_j+1} \frac{1}{\sqrt{2}}\big(\ket{\uparrow}_i\bra{x}_i-i\ket{\uparrow}\bra{y}_i-\ket{\downarrow}_i\bra{z}_i\big)\notag\\
c_{i;\downarrow}&= \prod_{j<i}(-1)^{n_j+1} \frac{1}{\sqrt{2}}\big(-\ket{\downarrow}_i\bra{x}_i-i\ket{\downarrow}\bra{y}_i-\ket{\uparrow}_i\bra{z}_i\big)\notag\\
\end{align}
where $\prod_{j<i}(-1)^{n_j+1}$ is the Jordan-Wigner string to enforce fermionic statistics.

In terms of $c_{i;\sigma}$, $n_i=\sum_\sigma c^\dagger_{i;\sigma}c_{i;\sigma}$, $\vec{S}^d_i=\sum_{\sigma,\sigma'}c^\dagger_{i;\sigma} \vec{\sigma}_{\sigma \sigma'}c_{i;\sigma'}$ and $\vec{S}^s_i=\sum_{\sigma,\sigma'}c_{i;\sigma'} \vec{\sigma}_{\sigma\sigma'} c^\dagger_{i;\sigma}$. Meanwhile  $c_{i;\sigma}c^\dagger_{i;\sigma}+c^\dagger_{i;\sigma}c_{i;\sigma}=1$ and $\{c_{i;\uparrow},c^\dagger_{i;\downarrow}\}=0$ do not hold anymore. Thus one should be careful in treating $c_{i;\sigma}$ as conventional electron operator. Anti-commutation relation between two operators of different sites still hold.

With the above definition of Hilbert space and physical operators, the type II $t-J$ model can be written as

\begin{align}
H_{t-J}&=H_t+H_J\\
H_t &=  - \sum_{\langle ij \rangle}t_{ij}c^\dagger_{i\sigma} c_{j\sigma} +{\rm h.c.}\\
H_J &= \sum_{\langle ij \rangle}\big( J  { \bf S^s_i \cdot S^s_j} + J_d {\bf S^d_i \cdot S^d_j} + \frac{J'}{2}({\bf S^s_i\cdot S^d_j +S^d_i\cdot S^s_j}) \notag\\
	&-(J_d+\frac{1}{4}J-J')n_i n_j\big)
\label{eq:t_J_model_main}
\end{align}

Generically we expect $J'\gtrsim J$.  Microscopically $t_{ij}$ is the hopping of the $d_{z^2}$ orbital and thus it has a large value in the $z$ direction\footnote{For square planar symmetry, it may also be possible that $d_{xy}$ orbital has lower energy than $d_{z^2}$ orbital in hole picture. In this case the hopping in $z$ direction should be negligible.}.  This means that the $t-J$ model for nickelate may need to be viewed as a three-dimensional model. Here, for simplicity, we will study the 2D version of the $t-J$ model and leave a 3D theory to future work.

We need to emphasize that the $c_{i\sigma}$ operator in Eq.~\ref{eq:t_J_model_main} is defined in the restricted Hilbert space. Even if we set $J=J'=0$, Eq.~\ref{eq:t_J_model_main} does not reduce to a free fermion model. $c_{i\sigma}$ annihilates a doublon state and creates a singlon at the same site. The hopping term in the $t-J$ model is essentially an exchange of singlon and doublon, which are therefre better variables than the electron operator electron operator $c_{i\sigma}$, to describe the underlying phases. At the limit $J,J'\rightarrow 0$, we expect a ferromagnetic ground state through the double exchange mechanism\cite{zener1951interaction}. However, superexchange terms $J,J'$ should suppress the ferromagnetism above a critical value. We will focus on the region that the $J,J'$ are large enough to favor a paramagnetic or anti-ferromagnetic ground state.

Intuitively there are two possible pictures in this novel $t-J$ model.  (I) In the simple picture, we just assume singlon-doublon separation. We can treat the doublon as a spin-one boson and it naturally condenses at finite density $x$. With the condensation of doublon, fermionic singlons can move coherently and form  Fermi surfaces and then  pair because of local anti-ferromagnetic spin coupling $J$. This is a simple generalization of the RVB theory\cite{anderson1987resonating,lee2006doping}.  However, in our case the condensation of the spin-one doublon necessarily breaks spin rotation symmetry. We call the resulting phases spin nematic d wave superconductor (SN-dSC) and spin-nematic Fermi liquid (SN-FL).  (II) In our $t-J$ model there is spin coupling $J'$ between the singlon and the doublon. Hence the singlon-doublon separation assumption may not be valid.  "Kondo resonance"  between singlons and doublons can be induced by $J'$. One can imagine a "heavy Fermi liquid" phase that each spin-one doublon contributes two particles and forms large Fermi surfaces together with the singlons, similar to "Kondo screening" in heavy fermion systems.  Because the doublon carries spin one, "Kondo screening" from the spin $1/2$ singlon happen in two steps. In the first stage, below a larger temperature $T^1_K$, "half" of the doublon is screened by the singlons, while the other half  forms a small hole pocket. In certain sense the physics can be understood in the following intuitive way: the doped hole enters the $d_{z^2}$ orbital and forms small pocket while there is a local spin $1/2$ moment sitting on $d_{x^2-y^2}$ orbital at every site. The resulting phase is either a fractionalized Fermi liquid (FL*) or an Antiferromagnetic ordered Fermi liquid with small Fermi surfaces.   Then in the second stage,  the small pocket absorbs the local spin $1/2$ to form a large Fermi surface below a Kondo scale $T^2_K$. Because of spin coupling $J$, the large Fermi surface gives way to a d-wave superconductor at lower temperature.

 In this paper we will show that this "Kondo resonance" picture can indeed naturally emerge in our $t-J$ model and be described by a novel  parton mean field theory. The parton theory can also describe the "SN-SC" and "SN-FL" phase from doublon condensation. Therefore we can study both scenarios above within one unified framework. Our mean field calculation shows that the "Kondo resonance" scenarios  wins unless there is a large external spin rotation breaking anisotropy.    In the following we first give a brief discussion of the SN-dSC phase through doublon condensation picture. Then we propose our new parton theory to describe both doublon condensation and Kondo resonance phases. A phase diagram based on mean field calculation will be provided.

\textit{\bf Slave boson theory} We introduce spin-one slave boson to label the triplet doublon, extending the popular slave boson theory of correlated electrons where the charge carrying boson is a spin singlet\cite{lee2006doping}.

\begin{align}
	c_{\uparrow}&=\frac{1}{\sqrt{2}}f^\dagger_{\uparrow}(b_x-ib_y)-\frac{1}{\sqrt{2}}f^\dagger_{\downarrow}b_z\notag\\
	c_{\downarrow}&=-\frac{1}{\sqrt{2}}f^\dagger_{\downarrow}(b_x+ib_y)-\frac{1}{\sqrt{2}}f^\dagger_{\uparrow}b_z\notag\\
	\label{eq:slave_boson}
\end{align}

$\vec{b}=(b_x,b_y,b_z)$ transforms as a vector under spin $SO(3)$ rotation. Spin operator of the doublon can be written as
\begin{equation}
    \vec S^d_{i}=-i \vec{b}^\dagger _i \times \vec{b}_i
\end{equation}

The constraint is $n_{b;i}+n_{f;i}=1$ and $n_{b;i},n_{f;i}=0,1$.  On average we have $\langle n_{b} \rangle=x$ and $\langle n_f \rangle =1-x$.

In the mean field theory, boson $b$ and fermion $f$ decouple. At finite density $x$, the bosonic spin-one doublon condenses to a spin-rotation breaking "superfluid" (which, of course, does not immediately imply a physical superfluid, since the slave bosons also carry gauge charge). For example,  consider the simple ansatz with $\langle b_x \rangle =0$. It breaks spin rotation but preserves the time reversal symmetry. Eq.~\ref{eq:slave_boson} shows that $c_\sigma \sim f^\dagger_\sigma$ and $f$ can now be viewed as electron operator.  Depending on the ansatz for $f$, we can obtain {\em either} a $d$ wave superconductor or a  Fermi liquid, with broken SO(3) spin rotation symmetry. In the presence of spin orbit coupling, the crystal structure will need to be considered to determine if there is actually any lowering of symmetry.  The mean field theory assuming slave boson condensation is exactly the same as that of the conventional $t-J$ model\cite{lee2006doping} and one expects a d-wave superconducting dome at small $x$. However other condensates such as $\langle \vec{b}\rangle =\vec{\psi_1}+i\vec{\psi_2}$, where $\vec{\psi_1}\times \vec{\psi_2}\neq0$ will correspond to ordered magnetic moments that break time reversal symmetry. We leave it to future work to determine the details of the symmetry breaking.

\textit{\bf Three-fermion parton theory} The spin-one slave boson approach does not include the possibility of Kondo resonance and can only describe FL/SC with spin rotation breaking. In this theory, the singlon and doublon decouple.  However, because of the $J'$ term, we expect that the singlon and doublon couple with each other through "Kondo resonance". Obviously we need a framework which can get access to both "Kondo resonance" regime and "Kondo breaking down" regime. In this section we show that this is possible in a new parton construction. We introduce two spin $1/2$ fermion $\Psi_\sigma=(\psi_{1\sigma},\psi_{2\sigma})^T$ to label the doublon state.  We label the doublon states by $\ket{a}=-\frac{1}{2\sqrt{2}}\Psi^\dagger  \tau_y  \sigma_a \sigma_y (\Psi^\dagger)^T \ket{0}$, where $\tau$ is the Pauli matrix in orbital space.  In the restricted five dimensional Hilbert space at each site, 

\begin{align}
	c_{\uparrow}&=f^\dagger_\uparrow \psi_{1\uparrow}\psi_{2\uparrow} +\frac{1}{2} f^\dagger_{\downarrow}(\psi_{1\uparrow}\psi_{2\downarrow}+\psi_{1\downarrow}\psi_{2\uparrow})\notag\\
	c_{\downarrow}&=f^\dagger_\downarrow \psi_{1\downarrow}\psi_{2\downarrow}+\frac{1}{2} f^\dagger_{\uparrow}(\psi_{1\uparrow}\psi_{2\downarrow}+\psi_{1\downarrow}\psi_{2\uparrow})\notag\\
\end{align}
under the constraint $n_{f;i}+n_{\psi_1;i}=1$ and $\Psi_{i,\sigma}^\dagger \tau_a \Psi_{i,\sigma}=0$. The later one constrains $\psi_{i;1},\psi_{i;2}$ to form orbital singlet, spin triplet at each site.  Again $f$ and $\psi_1,\psi_2$ are hard-core fermions whose density at each site can only be zero or one.  On average we have $\langle n_f \rangle=1-x$ and $\langle \psi_1 \rangle=\langle \psi_2 \rangle =x$.

We can see that the original electron (hole) operator is now written as a combination of three fermionic parton operators. We dub this parton construction as "three-fermion parton". A similar construction has been proposed for $SU(4)$ Hubbard model at total filling $\nu_T=1+x$\cite{zhang2019spin}.  There is a $SU(2)$ gauge symmetry: $\Psi_\sigma \rightarrow U \Psi_\sigma$ for $U \in SU(2)$.  There is another $U(1)$ gauge symmetry shared by $f$ and $\psi$: $f_i\rightarrow f_i e^{i \alpha_i}$ and $\Psi_i \rightarrow \Psi_i e^{i  \frac{1}{2} \alpha_i}$.  We assign the physical change in the way that $\psi_1,\psi_2$ carries $1/2$ charge while $f$ is neutral.

The spin operator is standard:
\begin{equation}
	\vec S^s=\frac{1}{2} f^\dagger_\alpha \vec{\sigma}_{\alpha \beta} f_\beta
\end{equation}
and
\begin{equation}
	\vec S^d=\frac{1}{2} \sum_{a=1,2} \psi^\dagger_{a,\alpha} \vec{\sigma}_{\alpha \beta} \psi_{a,\beta}
\end{equation}

$H_J$ in the $t-J$ model can be written using the above expressions. We can also rewrite the hopping term $t_2$ using the three fermion operators.  It is of the form: $f^\dagger_j f_i (\psi^\dagger_{i;2}\psi^\dagger_{i;1})(\psi_{j;1}\psi_{j;2})$ (see Append.~\ref{append:Hamiltonian_parton} for more details).

We can have a mean field theory by decoupling the original Hamiltonian.
\begin{align}
H_M&=- t_f \sum_{\langle ij \rangle}  f^\dagger_{i\sigma}f_{j\sigma}+h.c. 
- t^{\psi}_{ab}\sum_{ab=1,2}\sum_{\langle ij \rangle}  \psi^\dagger_{i;a}\psi_{j;b}+h.c. \notag\\
&-\Phi_a \sum_{ \langle ij \rangle }  (f^\dagger_{i\sigma} \psi_{j;a\sigma}+\psi^\dagger_{i;a\sigma} f_{j\sigma})+h.c.-\Phi^0_{a} \sum_{i} (f^\dagger_i \psi_{i;a}+\psi^\dagger_{i;a}f_i)\notag\\
&-\mu_f\sum_i n^f_i -\mu_1 \sum_i n^{\psi}_{i;1}-\mu_2 \sum_i n^\psi_{i;2} -\mu_{x} \sum_i (\psi^\dagger_{i;1}\psi_{i;2}+h.c.)\notag\\
&+  \sum_{\langle ij \rangle} \Delta_{f;ij} (f^\dagger_{i;\uparrow}f^\dagger_{j;\downarrow}-f^\dagger_{i\downarrow}f^\dagger_{j;\uparrow})+h.c.\notag\\
&+  \sum_{\langle ij \rangle} \Delta_{f,\psi_a; ij} (f^\dagger_{i;\uparrow}\psi^\dagger_{j;a\downarrow}-f^\dagger_{i;\downarrow}\psi^\dagger_{j;a\uparrow})+h.c.\notag\\
&+\Delta_{t} \sum_{i} (\psi^\dagger_{i;1\uparrow} \psi^\dagger_{i;2\uparrow}+\psi^\dagger_{i;1\downarrow} \psi^\dagger_{i;2\downarrow})+h.c.
\label{eq:mean_field_main}
\end{align}

$\mu_f,\mu_1, \mu_2$ are introduced to fix the density $\langle n_f\rangle =1-x$ and $\langle n_{\psi_1} \rangle=\langle n_{\psi_2} \rangle=x$. Meanwhile we need $\mu_{x}$  to fix the constraint that $\Psi^\dagger_i \tau_{x,y} \Psi_i=0$.

We have two sets of Kondo-like couplings: (I) $\Phi^0_a$ is on-site and is from the hopping term while $\Phi_a$ is between two nearest neighbor sites and originates  from the $J'$ coupling. Using $SU(2)$ gauge invariance we can remove one of them. Here we choose to fix $\Phi^0_2=0$; (II) From $J'$ coupling we also decouple a pairing term $\Delta_{f\psi_a}$. $\Delta_{f\psi_a}$ encodes the "Kondo resonance" that the singlon $f$ and $\psi_a$ want to form Kondo singlet. As we show later, to form a Fermi liquid with large Fermi surface, we need $\Phi_1 \neq 0$ and $\Delta_{f \psi_2}\neq 0$.

 We also introduce  spin-singlet pairing terms  between $f$ and between $\Psi$. From our mean field calculation, we find that the spin singlet pairings are {\em favored} to be of d-wave form.  Meanwhile we allow for an on-site triplet pairing $\Delta_t$ for $\Psi$, which is decoupled from the hopping term.

  Depending on the competition between different order parameters, we can have different phases in this framework. Here we list the most relevant ones in the following:

\begin{itemize}

\item $\Phi_1\neq 0$ (or $\Phi^0_1\neq 0$) and  $\Delta_{f\psi_2} \neq 0$. We find this solution at small $x$. Both $\Phi_1$ and $\Delta_{f\psi_2}$ can be viewed as Kondo-coupling from the $J'$ term. As we argue below, the resulting phase is a Fermi liquid with large Fermi surfaces ({\bf FL}). If we further include $\Delta_f \neq 0$, we get a $d$-wave superconductor ({\bf dSC}).

\item  $\Phi_1\neq 0, \Phi^0_1 \neq 0$ while $\Delta_{f\psi_2}=\Phi_2=0$.  In this case  $f$ is only coupled to $\psi_1$. The $f$ and $\psi_1$ hybridize to form a band with total filling $n=1$ per site. $\psi_2$ can now be identified as the physical hole operator and forms a small hole pocket with carrier density $x$. Analysis of gauge field (see Append. \ref{append:different_IGG_three_fermion}) shows that $f,\psi_1$ is neutral and they form  $U(1)$ spin liquid with spinon Fermi surfaces or $Z_2$ spin liquid with Dirac nodal fermions depending on whether $\Delta_f= 0$ or not, respectively. The resulting phase is the so called {\bf FL*} phase, with a small hole pocket of a Fermi liquid coexisting with a  spin-liquid\cite{senthil2004weak}.  

\end{itemize}

 For a fixed set of order parameters in the mean field theory, we can write down a variational wave function:  $\ket{\Psi}=P \ket{\tilde \Psi}$, where $\ket{\tilde \Psi}$ is a slater-determinant fixed by the mean field theory. At each site, the operator $P$ projects to the five states specified by $f^\dagger_{i;\sigma}\ket{0}$ and $(\Psi^\dagger)^T \tau_y \sigma_a \Psi^\dagger \ket{0}$. The order parameters in the mean field theory should be determined by  minimizing the energy corresponding to the projected wave-function. 

\begin{figure}[H]
\centering
\includegraphics[width=0.45\textwidth]{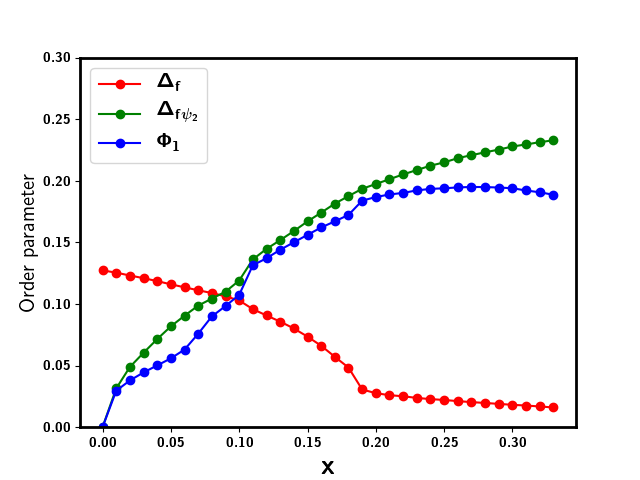}
\caption{Mean field solution from the three-fermion parton mean field theory using $t=2  J$ and $ J'=4  J$. We set $J=1$ and only show the dominant mean-field amplitudes.  }
\label{fig:phase_diagram_calculation}
\end{figure}

\begin{figure}[ht]
\centering
\includegraphics[width=0.45\textwidth]{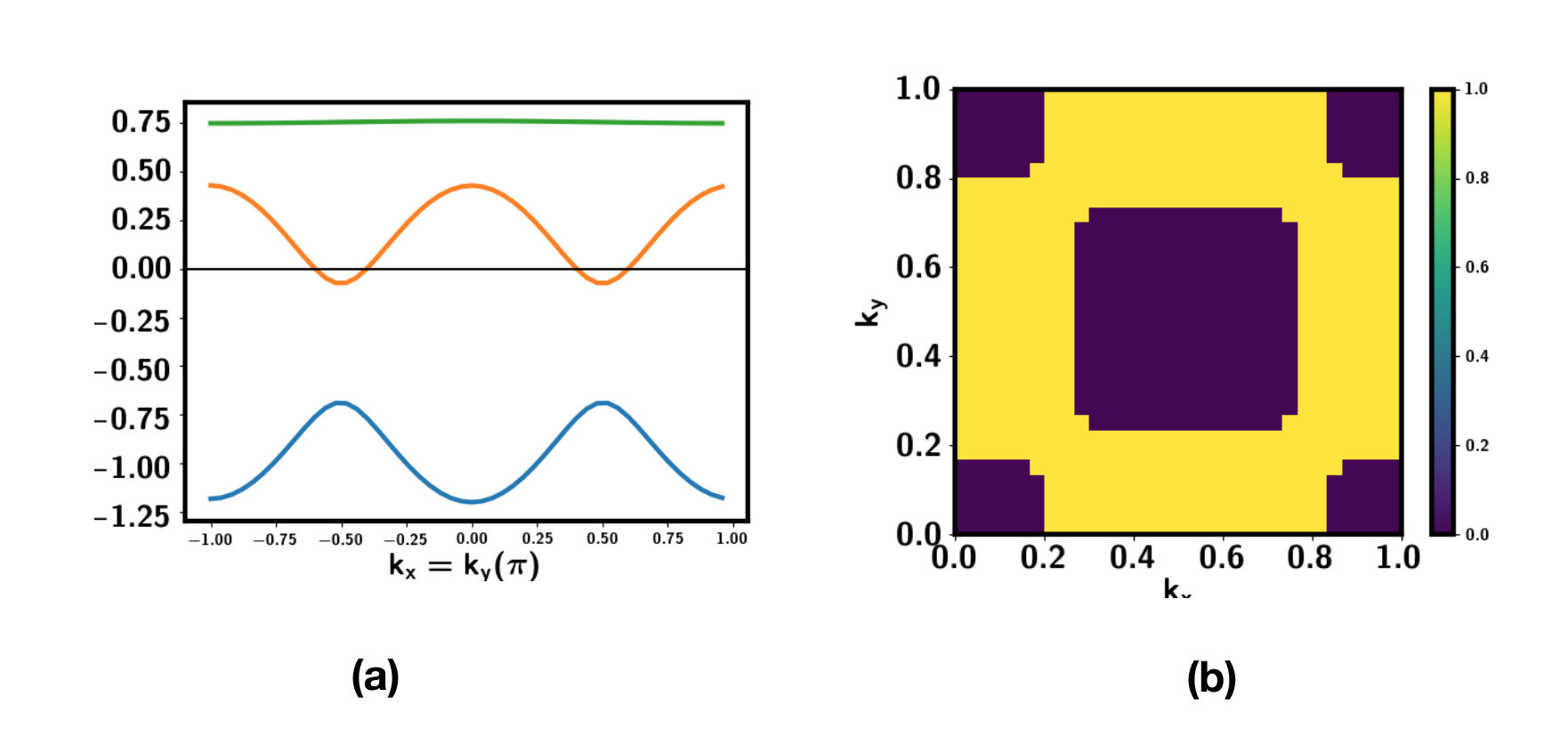}
\caption{Band structure at $x=20\%$ in terms of $\psi_{i;2\sigma}$, $\tilde f_{i;\sigma}=f^\dagger_{i;\sigma}$ and $\tilde \psi_{i;1\sigma}=\psi^\dagger_{i;1\sigma}$. We use $t=2J$ and $J'=4J$. Pairing term is suppressed by hand. For this specific choice of parameters, the resulting $n(\mathbf k)$ in Fig.(b) shows two electron pockets.}
\label{fig:band_structure}
\end{figure}

In the following we try to determine the order parameters in the level of mean field theory.  From solving the self consistent equations (shown in Append. ~\ref{appendix:mean_field_equation}), we get a plot of order parameters  shown in Fig.~\ref{fig:phase_diagram_calculation}.  The dominant order parameters are $\Phi_1$, $\Delta_{f\psi_2}$ and $\Delta_f$. First let us ignore $\Delta_f$.  We want to show that $\Phi_1$ and $\Delta_{f\psi_2}$ are Kondo couplings which merge $f,\psi_1, \psi_2$ to form a conventional Fermi liquid.   We have two $U(1)$ gauge field: $a$ is shared by $f$ and $\psi_a$, $\alpha$ parameterizes part of the $SU(2)$ gauge field  generated by $\tau_z$.  $f$ couples to $a$, $\psi_1$ couples to $\frac{1}{2}a +\frac{1}{2}\alpha+\frac{1}{2}A$. Now, $\psi_2$ couples to $\frac{1}{2}a-\frac{1}{2}\alpha+\frac{1}{2}A$.  The condensation of $\Phi_1$ and $\Delta_{f\psi_2}$ locks the gauge fields to be $a=-A$ and $\alpha=-2A$. Then $f$ and $\psi_1$ couple to $-A$  while $\psi_2$ couples to $A$.  This means that we can view $\psi_2$ as physical hole operator, while viewing $f,\psi_1$ as physical electron operator. Let us redefine $\tilde f_{i;\sigma}=f^\dagger_{i;\sigma}$ and $\tilde \psi_{i;1\sigma}=\psi^\dagger_{i;1\sigma}$. Then $\tilde f, \tilde \psi_1, \psi_2$ are all hole operators and they hybridize together to form a Fermi liquid with large Fermi surfaces, as shown in Fig.~\ref{fig:band_structure}.

At small $x$, we find $\Delta_f\neq 0$, thus the ground state is a d-wave superconductor. $\Delta_f$ decreases with doping, resulting a dome similar to that of the cuprates.  Here in the under-doped region, $T_c$ is decided by the on-set of $\Phi_1$ and $\Delta_{f\psi_2}$. In certain sense, the destruction of the superconductor is from "Kondo breaking down".

\begin{figure}[ht]
\centering
\includegraphics[width=0.45\textwidth]{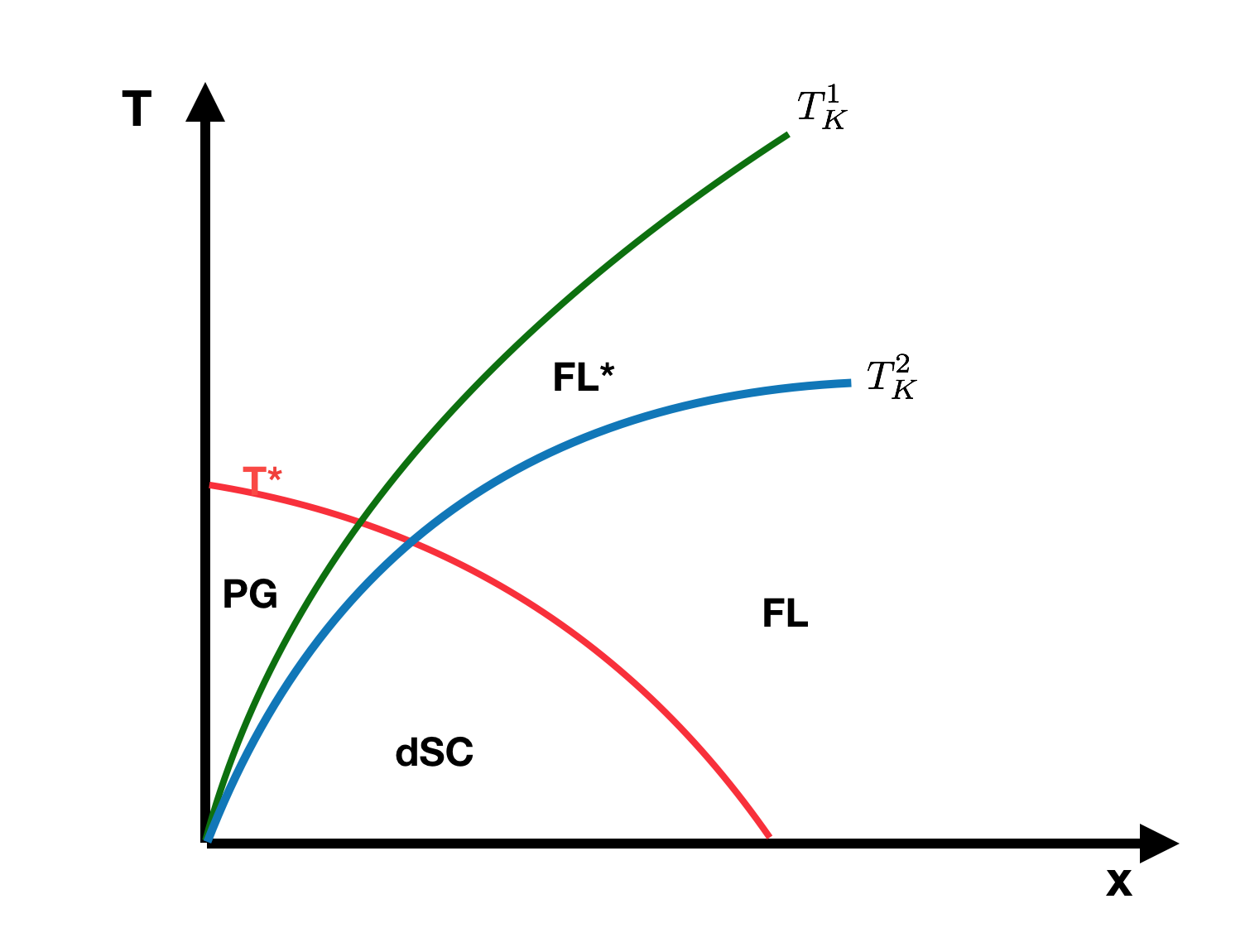}
\caption{Sketch of phase diagram in the $T-x$ space extrapolating from the zero temperature mean field theory. There are two "Kondo" scales: $T^1_K$ and $T^2_K$. $T^1_K$ is associated with $\Delta_{f\psi_2}$.  $T^2_K$ is associated with $\Phi_2$. $T^*$ determines the on-set of $\Delta_f$.}
\label{fig:phase_diagram_final}
\end{figure}

In the above we used $J'=4 J$ to get stable Fermi liquid/superconductor. For smaller $ J'/ J$, there is zero or just one Kondo coupling when $x<x_c$ in the mean field calculation, resulting in a  "pseudogap metals"  phase or a FL* phase in under-doped region.  It is not clear whether this is just an artifact of mean field treatment. Our mean field theory suggests that the phase diagram at small $x$ region is like that shown in Fig.~\ref{fig:phase_diagram_final}. One can see that this phase diagram is similar to that from slave boson theory for cuprates\cite{lee2006doping}. The difference is that here the $T_c$ (or coherence scale) is determined by "Kondo breaking down" instead of slave boson condensation.

\textit{\bf Role of Nd orbital} In the undoped sample, resistivity shows metallic behavior above $50$ K and an upturn below $50$ K. The metallic behavior can be attributed to electron pocket from Nd orbital\cite{Norman2019,ZXShen2019,Kuroki2019,Raghu2019}, leading to `self doping'.  We can extend our $t-J$ model to include the Nd orbital (see Appendix. ~\ref{append:Nd}):
\begin{align}
	H&=H_{t-J}+\sum_k \xi_{Nd}(\mathbf k)d^\dagger_\sigma(\mathbf k)d_\sigma(\mathbf k)\notag\\
	&+V \sum_i c^\dagger_i d_i+h.c.+J_{K} \vec S^s_i \cdot \vec S^{Nd}_i
\end{align}
where $d^\dagger_i$ creates a hole for Nd orbital.  Note here $c_i^\dagger$ creates the spin-one doublon and thus is different from the model in Ref.~\onlinecite{ZXShen2019}. If the density of Nd holes is $n_{Nd}=1-\delta$, then the density for spin-one holes is $n_c=x+\delta$.   In principle $\delta$ can have both doping and temperature dependence. Especially, $\delta$ is expected to decrease when we increase $x$\cite{Norman2019}. One can see that $n_c=\delta \neq 0$ even for $x=0$ if there is electron pocket from Nd in the Fermi level. This "self-doping" effect can explain why the undoped compound is metallic and does not have magnetic order.

\textit{\bf Conclusion} In summary, we propose a novel $t-J$ model with spin one hole (doublon in hole picture) to model the hole doped  NdNiO$_2$. We introduce two distinct parton theories to analyze this unconventional model. Especially, we find that a Fermi liquid or d wave superconductor is possible in this model, arising from the "Kondo resonance" between the spin one doublon and spin $1/2$ singly occupied state.  This suggests that a combination of "heavy fermion physics" and "cuprate physics" may emerge in this new model. We hope the proposed model and the parton framework introduced in this paper will motivate further theoretical and numerical studies, as well as experiments probing the spin-orbital nature of doped holes in this  new  superconductor.

\textit{ \bf Acknowledgement}
YH.Z thanks T. Senthil and Dan Mao for previous collaborations and discussions on $SU(4)$ Hubbard model and spin-one spin liquid, which inspired the proposal of the parton constructions in this paper. A. V. thanks Z. X. Shen and Julia Mundy for discussions. This work was supported by a Simons Investigator Grant (A.V.).

\bibliographystyle{apsrev4-1}
\bibliography{spin_one}

\onecolumngrid

\appendix

\section{Microscopic derivation of $t-J$ model with spin one doublon \label{append_derive_t_J}}

\subsection{Distinction from cuprates: spin one hole}

Because the oxygen $p$ orbital is far away from Fermi level in $NdNiO_2$, the doped hole will enter the $d$ orbitals.  From the LDA+U calculation \cite{Norman2019,ZXShen2019}, the energy splitting of the two $e_g$ orbitals is $0.7$ eV and smaller than the interaction scale. Therefore we need to include both $d_{x^2-y^2}$ and $d_{z^2}$ orbitals.  We choose to study the following model in hole picture:
\begin{align}
	H&=H_K+\frac{U_1}{2}\sum_i n_{1;i}(n_{1;i}-1)+\frac{U_2}{2}\sum_i n_{2;i}(n_{2;i}-1)+U'\sum_i n_{1;i}n_{2;i}-2J_H \sum_i (\mathbf{S}_{1;i}\cdot \mathbf{S}_{2;i}+\frac{1}{4}n_{i;1}n_{i;2})
\label{eq:spin_orbital_model}
\end{align}
where $n_{a;i}$ is the density of the orbital $a$ at the site $i$. $a=1,2$ denotes the $d_{x^2-y^2}$ and the $d_{z^z}$ orbital respectively. $U_1$, $U_2$ are intra-orbital Hubbard interaction. $U'$ is the inter-orbital interaction. $J_H$ is the inter-orbital Hund's coupling.  We expect $U_1=U_2=U$ and $U-U'=2J_H$ \cite{fresard1997interplay}.

The kinetic energy is
\begin{align}
	H_K&=\sum_i \epsilon_{dd} n_{2;i}+ \sum_{\langle ij \rangle} t_{1;ij} c^\dagger_{1;i}c_{1;j}+ \sum_{\langle ij \rangle} t_{2;ij} c^\dagger_{2;i}c_{2;j}+\sum_{\langle ij \rangle} t_{12;ij}c^\dagger_{1;i}c_{2;j}+h.c.
	\label{eq:full_Hamiltonian}
\end{align}
where $\epsilon_{dd}$ is the splitting between the two $e_g$ orbitals.

At zero doping, because $\epsilon_{dd}>0$, the ground state has one hole on the $d_{x^2-y^2}$ orbital at each site. Next we discuss the fate of the doped hole.  The energy cost for the hole to enter the orbital $1$ is $U_1$ while the energy cost for hole to be at orbital $2$ is $\epsilon_{dd}+U'-J_H$.   In the case that $\epsilon_{dd}<U-U'+J_H$, the orbital-singlet, spin triplet configuration is energetically favored. In this paper we take $U=3.4$ eV, $U'=2$ eV \cite{Kuroki2019} and $\epsilon_{dd}=J_H=0.7$ eV \cite{Norman2019,Kuroki2019}. We conclude that the doped hole in nickelate creates a spin one $d^8$ site. Therefore the low energy physics is governed by an unconventional $t-J$ model with spin-one hole.

\subsection{$t-J$ model}

Next we drive the low energy $t-J$ model.  First we need to define the Hilbert space. 

{\em Relation to the SU(4) Symmetric Model:} The model in Eq.~\ref{eq:spin_orbital_model} can be viewed as descending from a $SU(4)$ symmetric model but with  anisotropies that lower the symmetry. At the $SU(4)$ symmetric point, at the filling $\nu_T=1+x$, the $t-J$ model at the $U>>t$ limit has a Hilbert space with dimension $10=4+6$ at each site. These $10$ states can be divided to four singly occupied states  and six doubly occupied states\cite{zhang2019spin}. For simplicity let us call them singlon and doublon.  Singlon is in the fundamental representation of SU(4) while the doublon is in the $SO(6)$ representation\cite{zhang2019spin}.   In the nickelates, anisotropies can further constrain the Hilbert space to be five dimension at each site.  There are three large anisotropies: $\epsilon_{dd}$, $U-U'$ and $J_H$. 

We note here that the relevant hopping $t$ is entirely determined by $t_2$, the hopping between the $d_{z^2}$ orbitals. The hopping within the $d_{x^2-y^2}$ is blocked since we have eliminated doublons living in a single orbital ($U'\ll U$ limit), and inter orbital hopping is also eliminated for the same reason. We have not found an estimate for $t_2$, but we can estimate this to be of the order of $t_1 \sim 0.1$eV \cite{Kuroki2019}. Thus, we may assume we are in the limit $\epsilon_{dd}>>t$, $U-U'>>t$ and $J_H>>t$ limit, where we can further restrict the doublon states by doing the  $\frac{t}{U-U'}$ and $\frac{t}{J_H}$ expansion. 

First,  the $e_g$ orbital splitting $\epsilon_{dd}\sim 0.7$ eV\cite{Norman2019}. Therefore there are only two singlon states: $\ket{1\uparrow}$ and $\ket{1 \downarrow}$. The  $\ket{2\sigma}$ should not be included in the low energy Hilbert space.  However it does appear when we consider the doublons. For the six doublon states, because of the large $U-U'\sim 1-2$ eV, we should only consider the four states $\ket{1\sigma_1}\otimes  \ket{2\sigma_2}$. If $J_H<<t$, all of these four doublon states should be kept and we have $2+4=6$ states at each site.  However, in the opposite limit $J_H>>t$ which we assume, we should only include three doublon states corresponding to the three spin triplets. Therefore in total we only have $2+3=5$ states at each site in the low energy theory.

 \subsubsection{Labeling the Hilbert space}

 We focus on the $U, \epsilon_{dd}, U-U', J_H>>t$ limit and project to the five states at each site.  The Hilbert space at each site consists of a spin $1/2$ singlon and a spin one doublon.  We label the singlon with $\sigma=\uparrow, \downarrow$ and label the doublon with $a=x,y,z$. In terms of microscopic electron, $\ket{\sigma}=c^\dagger_{1\sigma} \ket{0}$ and 

\begin{align}
\ket{x} &=-\frac{1}{\sqrt{2}}(c^\dagger_{1\uparrow}c^\dagger_{2\uparrow}-c^\dagger_{1\downarrow}c^\dagger_{2\downarrow}) \ket{0}\notag\\
\ket{y} &=\frac{i}{\sqrt{2}}(c^\dagger_{1\uparrow}c^\dagger_{2\uparrow}+c^\dagger_{1\downarrow}c^\dagger_{2\downarrow}) \ket{0}\notag\\
\ket{z} &=\frac{1}{\sqrt{2}}(c^\dagger_{1\uparrow}c^\dagger_{2\downarrow}+c^\dagger_{1\downarrow}c^\dagger_{2\uparrow}) \ket{0}\notag\\
\label{eq:doublon_states_definition}
\end{align}

We also define a density operator at each site: $n_i= \sum_a \ket{a}\bra{a}$. $n_i$ measures the number of doublons.  In this restricted Hilbert space, the matrix elements of $c_{1\sigma}$ vanish, i.e. $c_{1\sigma}=0$. Also,  $c_{2\sigma}$ has the following matrix elements:

\begin{align}
c_{i;2\uparrow}&= \prod_{j<i}(-1)^{n_j} \frac{1}{\sqrt{2}}\big(\ket{\uparrow}_i\bra{x}_i-i \ket{\uparrow}\bra{y}_i-\ket{\downarrow}_i\bra{z}_i\big)\notag\\
c_{i;2\downarrow}&= \prod_{j<i}(-1)^{n_j} \frac{1}{\sqrt{2}}\big(-\ket{\downarrow}_i\bra{x}_i-i \ket{\downarrow}\bra{y}_i-\ket{\uparrow}_i\bra{z}_i\big)\notag\\
\end{align}
where $\prod_{j<i}(-1)^{n_j}$ is the Jordan-Wigner string to enforce fermionic statistics.

We can define spin operator $\mathbf{S^s}$ for the spin $1/2$ singlon and spin operator $\mathbf{S^d}$ for the spin-one doublon. It is easy to show that
\begin{align}
S^s_a=\frac{1}{2}  \sum_{\sigma,\sigma'} \sigma^{a}_{\sigma,\sigma'} \ket{\sigma}\bra{\sigma'}\notag\\
S^d_a=-i \sum_{b,c}\epsilon_{abc} \ket{b}\bra{c}
\end{align}
where Pauli matrices $\sigma$ and anti-symmetric tensor $\epsilon$ are used.

\subsubsection{Hamiltonian}

The $t-J$ Hamiltonian can be written as
\begin{equation}
	H=t_2 \sum_{ij} c^\dagger_{i;2\sigma}c_{j;2\sigma}+h.c.+H_J
\end{equation}
We need to emphasize that the Hamiltonian is defined in the restricted five states Hilbert space. Therefore $c_{i;2\sigma}$ shouldn't be confused with a conventional electron operator. In another word, $H_J=0$ does not reduce the Hamiltonian to a free fermion model.

 We should include super-exchange terms involving virtual hopping, which lead to the following spin coupling  
\begin{equation}
	H_J=J\sum_{\langle ij \rangle}(\vec{S^s_i}\cdot \vec{S^s_j}-\frac{1}{4}n^s_i n^s_j)+J_d\sum_{\langle ij \rangle}(\vec{S^d_i}\cdot \vec{S^d_j}-n^d_in^d_j)+\frac{1}{2}J'\sum_{\langle ij \rangle}\big((\vec{S^s_i}\cdot \vec{S^d_j}-\frac{1}{2}n^s_in^d_j)+(\vec{S^d_i}\cdot \vec{S^s_j}-\frac{1}{2}n^d_i n^s_j)\big)
\end{equation}
where $s,d$ label the spin operator for singlon and doublon.  $n^s_i=1-n_i$ and $n^d_i=n_i$ are the density of singlon and doublon states. 

 The spin coupling parameters are obtained from standard second order perturbation theory. We have  $J=4 \frac{t_1^2}{U}$, $J_d=\frac{t_1^2}{U}+\frac{t_2^2}{U}+2\frac{t_{12}^2}{U}$. $J'=\frac{1}{2}(J_1+J_2)+\frac{t_2^2}{2J_H}-t_{12}^2(\frac{1}{\epsilon_{dd}}-\frac{1}{\epsilon_{dd}+J_H})$ where $J_1=2(\frac{t_1^2}{U_1+U'}+\frac{t_1^2}{U_1-U'})$ and $J_2=2(\frac{t_{12}^2}{U_2+U'+\epsilon_{dd}}+\frac{t_{12}^2}{U_1-U'-\epsilon_{dd}})$. $J'$ term contains a contribution proportional to $\frac{t_2^2}{2J_H}$, which is from integrating the orbital-triplet, spin singlet doublon. The final value of $J'$ depends on the details of the material, but it is definitely reasonable to assume that $J'>J$.  Actually if we assume $U=2 U'=4J_H$, $\epsilon_{dd}=J_H$, $t_1=t_{12}$ and $t_2=0$ ($t_2$ should be smaller in $xy$ plane because it is associated with the $d_{z^2}$ orbital), we find $J' \approx 1.3 J$.  In the following we will view $\frac{J'}{J}$ as a phenomenological parameter. The $J'$ term can be viewed as a Kondo coupling between the spin $1/2$ singlon and spin $1$ doublon. It can cause "Kondo resonance" between them.

\section{Hamiltonian in the spin-one slave boson and three-fermion parton theory \label{append:Hamiltonian_parton}}

In the main-text we introduced two different parton constructions: spin-one slave boson and a three-fermion parton. Here we show that there is a connection between these two parton theories. The three-fermion parton can be derived from the slave boson parton by further fractionalizing the spin-one slave boson to two spin $1/2$ fermions:

\begin{align}
b_x^\dagger &=-\frac{1}{\sqrt{2}}(\psi^\dagger_{1\uparrow}\psi^\dagger_{2\uparrow}-\psi^\dagger_{1\downarrow}\psi^\dagger_{2\downarrow})\notag\\
b_y^\dagger &=\frac{i}{\sqrt{2}}(\psi^\dagger_{1\uparrow}\psi^\dagger_{2\uparrow}+\psi^\dagger_{1\downarrow}\psi^\dagger_{2\downarrow})\notag\\
b_z^\dagger &=\frac{1}{\sqrt{2}}(\psi^\dagger_{1\uparrow}\psi^\dagger_{2\downarrow}+\psi^\dagger_{1\downarrow}\psi^\dagger_{2\uparrow})\notag\\
\end{align}

The physical hole operator can then be written as

\begin{align}
	c_{\uparrow}&=\frac{1}{\sqrt{2}}f^\dagger_{\uparrow}(b_x-ib_y)-\frac{1}{\sqrt{2}}f^\dagger_{\downarrow}b_z=f^\dagger_\uparrow \psi_{1\uparrow}\psi_{2\uparrow} +\frac{1}{2} f^\dagger_{\downarrow}(\psi_{1\uparrow}\psi_{2\downarrow}+\psi_{1\downarrow}\psi_{2\uparrow})\notag\\
	c_{\downarrow}&=-\frac{1}{\sqrt{2}}f^\dagger_{\downarrow}(b_x+ib_y)-\frac{1}{\sqrt{2}}f^\dagger_{\uparrow}b_z=f^\dagger_\downarrow \psi_{1\downarrow}\psi_{2\downarrow}+\frac{1}{2} f^\dagger_{\uparrow}(\psi_{1\uparrow}\psi_{2\downarrow}+\psi_{1\downarrow}\psi_{2\uparrow})\notag\\
\end{align}

In terms of the slave boson construction, the hopping term in the $t-J$ model can be rewritten as

\begin{align}
	H_K&=-\frac{t_2}{2} \sum_{\langle ij \rangle}f^\dagger_{j\uparrow}f_{i\uparrow}(b^\dagger_{i;x}b_{j;x}+b^\dagger_{i;y}b_{j;y}+b^\dagger_{i;z}b_{j;z})\notag\\
	&-\frac{t_2}{2} \sum_{\langle ij \rangle}f^\dagger_{j\downarrow}f_{i\downarrow}(b^\dagger_{i;x}b_{j;x}+b^\dagger_{i;y}b_{j;y}+b^\dagger_{i;z}b_{j;z})\notag\\
	&-t_2 \sum_{\langle ij \rangle}\sum_{a=x,y,z} (f_j^\dagger S^f_a f_i)  (b^\dagger_i S^b_a b_j)
\end{align}

In terms of the three-fermion  parton construction, the kinetic part is

\begin{align}
	H_K&=-\frac{1}{4}t_2 \sum_\sigma \sum_{\langle ij \rangle}f^\dagger_{j\sigma}f_{i\sigma}(\psi^\dagger_{i;2\downarrow}\psi^\dagger_{i;1\uparrow}+\psi^\dagger_{i;2\uparrow}\psi^\dagger_{i;1\downarrow})(\psi_{j;1\uparrow}\psi_{j;2\downarrow}+\psi_{j;1\downarrow}\psi_{j;2\uparrow})\notag\\
	&-t_2 \sum_\sigma \sum_{\langle ij \rangle} f^\dagger_{j;\sigma}f_{i;\sigma} \psi^\dagger_{i;2\sigma}\psi^\dagger_{i;1\sigma}\psi_{j;1\sigma}\psi_{j;2\sigma}\notag\\
	&-\frac{1}{2} t_2 \sum_{\langle ij \rangle}f^\dagger_{j;\uparrow}f_{i;\downarrow}\left((\psi^\dagger_{i;2\downarrow}\psi^\dagger_{i;1\uparrow}+\psi^\dagger_{i;2\uparrow}\psi^\dagger_{i;1\downarrow})(\psi_{j;1\uparrow}\psi_{j;2\uparrow})+(\psi^\dagger_{i;2\downarrow}\psi^\dagger_{i;1\downarrow})(\psi_{j;1\uparrow}\psi_{j;2\downarrow}+\psi_{j;1\downarrow}\psi_{j;2\uparrow})\right)\notag\\
	&-\frac{1}{2} t_2 \sum_{\langle ij \rangle}f^\dagger_{j;\downarrow}f_{i;\uparrow}\left((\psi^\dagger_{i;2\downarrow}\psi^\dagger_{i;1\uparrow}+\psi^\dagger_{i;2\uparrow}\psi^\dagger_{i;1\downarrow})(\psi_{j;1\downarrow}\psi_{j;2\downarrow})+(\psi^\dagger_{i;2\uparrow}\psi^\dagger_{i;1\uparrow})(\psi_{j;1\uparrow}\psi_{j;2\downarrow}+\psi_{j;1\downarrow}\psi_{j;2\uparrow})\right)\notag\\
	&+h.c.
\end{align}

\section{Mean field theory and self-consistent equations \label{appendix:mean_field_equation}}

We list the mean field theory from the three-fermion parton and self-consistent equations for a very general mean field ansatz here.  We includes the spin-triplet pairing term of $\Psi$ in the mean field ansatz. Triplet pairing breaks the $SO(3)$ spin rotation, we choose the pairing channel to be $\psi_{i;1\uparrow}\psi_{i;2\uparrow}-\psi_{i;1\downarrow}\psi_{i;2\downarrow}$. This pairing has the same spin rotation symmetry as $b_x^\dagger$.

We can have a mean field theory by decoupling the original Hamiltonian.
\begin{align}
H_M&=- t_f \sum_{\langle ij \rangle}  f^\dagger_{i\sigma}f_{j\sigma}+h.c. 
- t^{\psi}_{ab}\sum_{ab=1,2}\sum_{\langle ij \rangle}  \psi^\dagger_{i;a}\psi_{j;b}+h.c. \notag\\
&-\Phi_a \sum_{ \langle ij \rangle }  (f^\dagger_{i\sigma} \psi_{j;a\sigma}+\psi^\dagger_{i;a\sigma} f_{j\sigma})+h.c.-\Phi^0_{a} \sum_{i} (f^\dagger_i \psi_{i;a}+\psi^\dagger_{i;a}f_i)\notag\\
&-\mu_f\sum_i n^f_i -\mu_1 \sum_i n^{\psi}_{i;1}-\mu_2 \sum_i n^\psi_{i;2} -\mu_{x} \sum_i (\psi^\dagger_{i;1}\psi_{i;2}+h.c.)\notag\\
&+  \sum_{\langle ij \rangle} \Delta_{f;ij} (f^\dagger_{i;\uparrow}f^\dagger_{j;\downarrow}-f^\dagger_{i\downarrow}f^\dagger_{j;\uparrow})+h.c.\notag\\
&+  \sum_{\langle ij \rangle} \Delta_{f,\psi_a; ij} \big((f^\dagger_{i;\uparrow}\psi^\dagger_{j;a\downarrow}-f^\dagger_{i;\downarrow}\psi^\dagger_{j;a\uparrow})+(\psi^\dagger_{i;a\uparrow}f^\dagger_{j;\downarrow}-\psi^\dagger_{i;a\downarrow}f^\dagger_{j;\uparrow}) \big)+h.c.\notag\\
&+\Delta_{t} \sum_{i} (\psi^\dagger_{i;1\uparrow} \psi^\dagger_{i;2\uparrow}+\psi^\dagger_{i;1\downarrow} \psi^\dagger_{i;2\downarrow})+h.c.
\label{eq:mean_field_append}
\end{align}

$\mu,\mu_1, \mu_2$ are introduced to fix the density $\langle n_f\rangle =1-x$ and $\langle n_{\psi_1} \rangle=\langle n_{\psi_2} \rangle=x$. Meanwhile we need $\mu_{12}$ and $\mu_{21}$ to fix the constraint that $\Psi^\dagger_i \tau_{x,y} \Psi_i=0$.

Although there is triplet pairing, there is still time reversal symmetry. Therefore any correlation for spin up and spin down is guaranteed to be equal. This can greatly simplify our self-consistent equations:

Following the standard variational principle\cite{brinckmann2001renormalized}, self-consistent equations can be derived:

\begin{align}
t^f&=\frac{3}{8}   J \chi_f  +\frac{3}{8}  t (\chi_1 \chi_2 -\chi_{21}\chi_{12})+\frac{1}{4}  t |\chi_{\Delta_t}|^2\notag\\
t^{\psi}_1&=\frac{3}{8}   t \chi_f \chi_2-\frac{9}{16}  t C_{2j}C_{2i} \notag\\
t^{\psi}_2&=\frac{3}{8}  t \chi_f \chi_1-\frac{9}{16}  t C_{1j}C_{1i} \notag\\
t^{\psi}_{12}&=-\frac{3}{8}   t \chi_f \chi_{21}+\frac{9}{16}t C_1 C_2 \notag\\
\Delta_{ij}&=\frac{3}{8}   J \chi_\Delta \notag\\
\Delta_t&=-\frac{1}{2}  t \chi_f \chi_{\Delta_t} \notag\\
\Delta_{f\psi_a}&=\frac{3}{16}   J' \chi_{\Delta_{f\psi_a}}\notag\\
\Phi_a &=\frac{3}{16}   J' C_a \notag\\
\Phi^0_1 &=  t  \sum_{j \sim i} (-\frac{5}{8} C_1 \chi_2 + \frac{5}{8} C_2 \chi_{12})\notag\\
\Phi^0_2 &=  t  \sum_{j \sim i} (-\frac{5}{8} C_2 \chi_1 + \frac{5}{8} C_1 \chi_{12})\notag\\
\end{align}

where
\begin{align}
\chi_f&=\sum_\sigma \langle f^\dagger_{j;\sigma} f_{i;\sigma} \rangle \notag\\
\chi^\psi_a&=\sum_\sigma \langle \psi^\dagger_{j;a\sigma} \psi_{i;a\sigma} \rangle \notag\\
\chi^\psi_{12}&=\sum_\sigma \langle \psi^\dagger_{j;1\sigma} \psi_{i;2\sigma} \rangle \notag\\
C^0_a&=\sum_\sigma \langle \psi^\dagger_{j;a\sigma} f_{j;\sigma}\rangle \notag\\
C_a&=\sum_\sigma \langle \psi^\dagger_{j;a\sigma} f_{i;\sigma}\rangle \notag\\
\chi_\Delta&=\langle f_{i\uparrow}f_{j\downarrow}-f_{i\downarrow}f_{j\uparrow}\rangle \notag\\
\chi_{\Delta_{f\psi_a}}&=\langle \psi_{i;a\uparrow}f_{j\downarrow}-\psi_{i;a\downarrow}f_{j\uparrow}\rangle \notag\\
\chi_{\Delta_t}&=\langle \psi_{i;2\uparrow}\psi_{i;1\uparrow}+\psi_{i;2\downarrow}\psi_{i;1\downarrow}\rangle \notag\\
\end{align}

\begin{figure}
    \centering
    \includegraphics[width=0.9\textwidth]{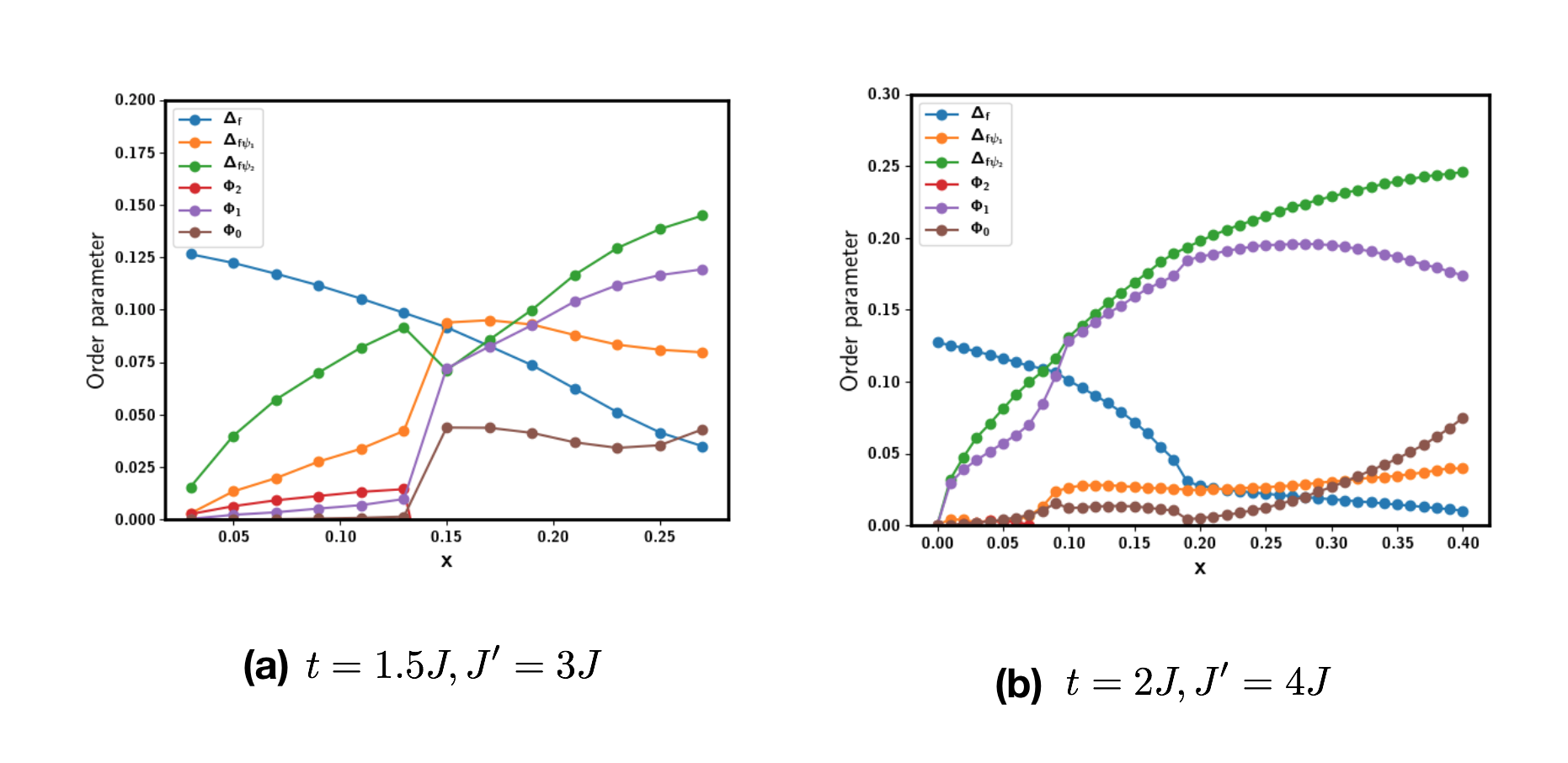}
    \caption{Order parameters from mean field equations.}
    \label{fig:order_plot}
\end{figure}

The mean field solutions are shown in Fig.~\ref{fig:order_plot}. In our mean field calculation, the Kondo-couplings $\Phi_1$ and $\Delta_{f\psi_2}$ decreases rapidly when decreasing $J'$. It is not clear whether this is an artifact of our naive mean field treatment or not.

\section{$J_H \rightarrow 0$ limit: heavy fermion physics\label{append_JH_0}}

As we argued previously, a proper model for nickelate is a $t-J$ model with a spin-one doublon, which is derived from $J_H>>t$ limit. In this section we show that a heavy fermion like model can be written down at finite $J_H$. The model then reduces to the $t-J$ model with spin-one doublon if we take $J_H\rightarrow \infty$ limit. As we will see later, this model at small $J_H$ actually captures some of the essential physics connecting to the $J_H>>t$ limit.

At finite $J_H$, we should keep $2+4=6$ states at each site. At each site the $6=2\times 3$ dimensional Hilbert space has a tensor product structure: $H_i=H^1_i \otimes H^2_i$ where $H^a_i$ means the Hilbert space for orbital $a=1,2$ at site $i$. $H^1_i$ is a spin $1/2$ on orbital one and $H^2_i$ is a three dimensional space generated by the hole operator $c^\dagger_{2\sigma}$ with the constraint $n_{i;2}=0,1$.   Basically, the doped hole enters the orbital $2$ while the density on orbital one is fixed to be one per site.  Therefore, the final theory consists of $xN_s$ $c_{2\sigma}$ particles moving and interacting with spin $1/2$ moment at each site.

The final model is quite similar to a Kondo-Heisenberg lattice model:

\begin{align}
H&=t_2 \sum_{\langle ij \rangle} P c^\dagger_{i;2\sigma}c_{j;2\sigma} P+h.c.-\mu \sum_i c^\dagger_{i;2\sigma}c_{i;2\sigma}\notag\\
&- 2 J_H \sum_i \vec{S^{2}_i} \cdot \vec{S^1_i} + J \sum_{\langle ij \rangle} \vec{S^1_i}\cdot \vec{S^1_j}\notag\\
&+\frac{1}{2}(J_1-J)\sum_{\langle ij \rangle} \left(n_{i;2}(1-n_{j;2})+(1-n_{i;2})n_{j;2}\right)\vec{S^1_i}\cdot \vec{S^1_j}\notag\\
&+\frac{1}{2}J_2 \sum_{\langle ij \rangle}(\vec{S^1_i}\cdot \vec{S^2_j}+\vec{S^2_i}\cdot \vec{S^1_j})+J_d \sum_{\langle ij \rangle} \vec{S^2_i}\cdot \vec{S^2_j}
\label{eq:JH_0_t_J}
\end{align}
where $J_1,J_2$ is the spin coupling between the singlon and the spin of the doublon. $(J_1-J)$ term reflects the fact that the coupling between two nearest neighbor spin on orbital $1$ is different if one site is a singlon and the other one is a doublon.  $P$ constrains $n_{i;2}=0,1$.  On average we have $\langle n_{i;2} \rangle=x$.  

We have $J=4 \frac{t_1^2}{U_1}$, $J_d=4 \frac{t_2^2}{U_2}$.  $J_1=2(\frac{t_1^2}{U_1+U'}+\frac{t_1^2}{U_1-U'})$ and $J_2=2(\frac{t_{12}^2}{U_2+U'+\epsilon_{dd}}+\frac{t_{12}^2}{U_1-U'-\epsilon_{dd}})$.  Assuming $U_1=U_2=U$ and $U'=\frac{1}{2}U$ and $\epsilon_{dd}=\frac{1}{4}U$, we have $J_1 = \frac{4}{3} J$.  If we further assume $t_{12}=t_1$, then $J_2=\frac{4}{3}J$. 

In the following we assume $J_1=J$ for simplicity.  Then the above model resemble the models for heavy fermion systems. Basically the particle on orbital $1$ contributes a local spin $1/2$ moment while oribital $2$ provide itinerant particles. In this model, the itinerant electrons couple to the local moment through both ferromagnetic $J_H$ and anti-ferromagnetic "Kondo" coupling.  The model in Eq.~\ref{eq:JH_0_t_J} reduces to the model in Eq.~\ref{eq:t_J_model_main} if we take the $J_H \rightarrow \infty$ limit. (Interestingly, the $J_H\rightarrow -\infty$ limit gives the conventional $t-J$ model in cuprate. )

When $J_H=0$, itinerant holes from $c_{2\sigma}$ couples to the spin $1/2$ moment. "Heavy fermi liquid" or heavy fermion superconductor may show up through Kondo resonance.  However, the large $J_H$ limit is not clear at all from the above Hamiltonian. With a large $J_H$, $c_{2\sigma}$ itself can not move coherently because it strongly couples to the spin $1/2$ at the same site. Then the $t-J$ model in Eq.~\ref{eq:t_J_model_main} may be a better starting point. However, as we show in the main text, even in the $J_H\rightarrow \infty$ limit, similar "heavy fermion" physical may emerge in the low energy. But there the small hole pocket is formed by fermion $\psi_{2\sigma}$ below $T^1_K$ (see Fig.~\ref{fig:phase_diagram_final}), which is not identical to the microscopic $c_{2\sigma}$. We may view this $\psi_{2\sigma}$ operator as a "renormalized" operator in the infrared limit of RG flow. Our mean field theory of the $t-J$ model in the $J_H\rightarrow \infty$ limit shows that this "renormalized hole" $\psi_{2\sigma}$ can move coherently and does not feel frustration from $J_H$ (after all $J_H$ disappears in the $t-J$ model.)   Depending on whether there is Kondo resonance between $\psi_{2\sigma}$ and the spin $1/2$ moment, we have a Fermi liquid with large Fermi surface (or superconductor) or a fractionalized Fermi liquid (FL*) phase.

\section{Different Phases in the three-fermion parton theory \label{append:different_IGG_three_fermion}}

We discuss different phases described by the mean field theory in Eq.~\ref{eq:mean_field_append} in the three-fermion parton theory. For simplicity, we focus on the ansatz with $\Delta_f=0$. The pairing term can be added later. Our parton construction has a $U(1)\times SU(2)$ gauge invariance. We can always choose a $SU(2)$ gauge to remove $f^\dagger_i \psi_{i;2}$ term.  We can have different phases corresponding to remaining gauge structure (invariant gauge group (IGG)).

\subsubsection{$U(1)\times SU(2)$: Kondo breaking down}

 If $\Phi^0_1=\Phi_1=\Phi_2=\Delta_{f\psi_a}=0$ and $t^\psi_1=t^\psi_2$, $t^\psi_{12}=0$, our mean field ansatz still has the full $U(1)\times SU(2)$ structure. In this case $f$ and $\psi$ are decoupled.  From the decoupling of  $H_K$, $\psi$ can get both  hopping and triplet-pairing term. Let us forget the triplet-pairing term first.  Then $\psi_1$ and $\psi_2$ form two separate hole pockets, which couple to $U(1)$ gauge field $a$ and $SU(2)$ gauge field $\alpha$.  $SU(2)$ gauge field  always mediates attractive interaction in the orbital singlet, spin triplet channel.   Therefore we conclude that there is no stable phase with the IGG $U(1)\times SU(2)$.  $\Psi$ is always gapped out by a spin-triplet pairing term $\Delta_t$  and then $SU(2)$ gauge field is confined.  In this case, the  resulting phase  is exactly the same as that accessed by condensing spin-one slave boson in the previous section.  Depending on the ansatz of $f$, we can  have  either spin-nematic $d$ wave superconductor or spin-nematic Fermi liquid. 

\subsubsection{$U(1)_a \times U(1)_\alpha$: $U(1)$ pseudogap metal }

In this IGG, we still need $\Phi^0_1=\Phi_2=\Phi_0=\Delta_{f\psi_a}=0$. However, we introduce $t^\psi_1 \neq t^\psi_2$ to  higgs $SU(2)$ to $U(1)$.  $U(1)$ is generated by $\tau_z$ and let us label it as $\alpha$. Meanwhile there is another $U(1)$ gauge field $a$ shared by $f$ and $\Psi$. They have the following charges - $f$ couples to $a$, $\psi_1$ couples to $\frac{1}{2}A+\frac{1}{2}a+\frac{1}{2}\alpha$ and  $\psi_2$ couples to $\frac{1}{2}A+\frac{1}{2}a-\frac{1}{2}\alpha$. 

If $\Delta_f=0$, we have a large Fermi surface from $f$ and two smaller Fermi surfaces from $\psi$.  This is a very exotic metal similar to the "deconfined metal" proposed in Ref.~\onlinecite{zhang2019spin} for the $SU(4)$ model. 

If $\Delta_f \neq 0$, $a$ is Higgsed down to $Z_2$ and we can ignore it. We still have two small Fermi surfaces formed by $\psi_1$ and $\psi_2$. Meanwhile $\psi_1$ couples to $\frac{1}{2}A+\frac{1}{2}\alpha$ while $\psi_2$ couples to $\frac{1}{2}A-\frac{1}{2}\alpha$.  Because $f$ forms $d$ wave pairing, physical electron must have anti-node gap. Around node $(\frac{\pi}{2},\frac{\pi}{2})$ there should still be gapless excitations from convolution of $f,\psi_1,\psi_2$.  We dub this phase as " $U(1)$ pseudogap metal" because of its similarity to the pseudogap metal in cuprates and a deconfined $U(1)$ gauge field.

\subsubsection{$U(1)_a$: deconfined metal and $Z_4$ pseudogap metal}
We assume ansatz $t^\psi_1 \neq t^\psi_2$ and $t^\psi_{12} \neq 0$ for $\Psi$, which fully higgses the $SU(2)$ gauge field. Besides we need $\Phi_0=\Phi_1=\Phi_2=\Delta_{f\psi_a}=0$.  Then we only have $U(1)$ gauge field $a$.  In this case, we have two sets of Fermi surfaces formed by $f$ and $\psi_{1,2}$.   The low energy theory looks like:

\begin{equation}
	L=L_{FS}[f,2a]+L_{FS}[\psi,\frac{1}{2}A+a]
\end{equation}
We call this phase as "deconfined metal" because of a deconfined $U(1)$ gauge field. The areas of the two Fermi surfaces are $1-x$ and $x$ respectively.  They are strongly coupled together by gauge field $a$.

When there is also pairing for $f$: $\Delta_f\neq0$ which higgses $a$ down to $Z_4$,  we get a "pseudogap metal". But in this case the Fermi surface formed by $\psi$ only couples to a $Z_4$ gauge field.  There may be one or two hole pockets depending on details of hopping terms in $\psi$. But generically this is a stable metallic phase. Its property is similar to "orthogonal metal"\cite{nandkishore2012orthogonal} except that now the $Z_4$ gauge theory part also contains gapless nodal fermion from $f$.  

This "psuedogap phase" has the following properties: (I) The thermodynamic properties and transport properties are the same as Fermi liquid with small hole pockets.  The size of the hole pocket is equal to $2x$. But the physical charge carried by $\psi$ is only $1/2$. Therefore Hall number is $x$.   (II) Green function of $c$ is a convolution of $f, \psi_1, \psi_2$.  $f$ only has gapless excitation around $(\frac{\pi}{2},\frac{\pi}{2})$. $\psi_1,\psi_2$ is likely to have pockets at either $(0,0)$ or $(\pi,\pi)$.  The spectral function of physical electron operator must only have gapless weights at around $(\frac{\pi}{2},\frac{\pi}{2})$ after convolution.  Both of these features resemble the pseudogap metal in hole doped cuprate.

\subsubsection{$U(1)_\alpha$: FL* }

We add $\Phi_1 \neq 0$ which hybridizes $f$ and $\psi_1$ but still assume $\Phi_2 = \Phi_0=0$ and $t^\psi_{12}=0$.  Therefore $\psi_2$ is decoupled from $f,\psi_1$.

$f$ couples to $2a$. $\psi_1$ couples to $\frac{1}{2}A+a+\frac{1}{2}\alpha$. $\psi_2$ couples to $\frac{1}{2}A+a-\frac{1}{2}\alpha$.  After $\Phi \neq 0$, we have $a=\frac{1}{2}(A+\alpha)$.  Therefore the final phase still has one deconfined $U(1)$ gauge field.

There are two sets of Fermi surfaces in the low energy. The first one is formed by $f,\psi_1$ and it couples to $A+\alpha$. The other one is formed by $\psi_2$ and it couples to $A$.   We have $c\sim f^\dagger \psi_1\psi_2$. After adding $f^\dagger \psi_1$, $\psi_2$ is the same as physical electron.  We can always redefined $\tilde \alpha=A+\alpha$. The final theory is

\begin{equation}
	L= L_{FS}[f,\tilde \alpha]+L_{FS}[\psi_2,A]
\end{equation}

This describes a FL* phase. Basically the Fermi liquid part from $\psi_2$ coexists with a neutral fermi surface formed by $f,\psi_1$. Because there is no coupling like $\psi^\dagger_1 \psi_2$, the two Fermi surfaces do not merge.   Our constraint is $n_{\psi_1}=n_{\psi_2}=x$ and $n_f=1-x$.  Thus the Fermi surface area of the Fermi liquid part is fixed to be $x$.  Naively $f$ forms a $U(1)$ spin liquid with spinon Fermi surface and the Fermi surface area is $1/2$. In practice the spin liquid  part may likely be confined. Then the phase is just a small hole pocket decoupled from spin $1/2$ moment. It is exactly the "small Fermi surface" phase above Kondo scale we descirbed in the $J_H=0$ limit.

Next we add pairing  $\Delta_f$ to the mean field ansatz.  The neutral part becomes a $Z_2$  Dirac spin liquid and we still call the resulting phase FL*. 

\subsubsection{Fully higgsed: conventional Fermi liquid with large Fermi surface}

Consider ansatz with $\Phi_0\neq 0$ and $\Delta_{f\psi_2} \neq 0$. All of gauge fields are fully higgsed.  We finally have a conventional Fermi liquid.  The fermi surface area is decided by $n_f+n_1+n_2=1+x$, consistent with the Luttinger theorem.  If we further introduce pairing, we get a conventional superconductor.

\section{Role of $Nd$ orbitals \label{append:Nd}}
Here comment on the possible role of $5d$ orbitals of the $Nd$ element. Ref.~\onlinecite{ZXShen2019} suggests that the $Nd$ orbital couples to the $d_{x^2-y^2}$ orbital of $N_i$ like in Anderson model, which gives Kondo resonance.  At zero doping, there is indeed fluctuation between $d^9$ state and $d^8 R$ state where $R$ denotes the $Nd$ orbital. However, as we argued previously, the lowest energy $d^8$ state of Ni is a spin triplet occupying both $e_g$ orbitals. Thus a more appropriate lattice model should involve both $e_g$ orbitals of Ni and the dominant process is to create the spin-one $d^8$ state:
\begin{align}
	H&=\sum_{k}\epsilon^1_k c^\dagger_1(\mathbf k) c_1(\mathbf k)+\sum_{k}(\epsilon^2_k+\epsilon_{dd}) c^\dagger_2(\mathbf k) c_2(\mathbf k)+\sum_{\mathbf k} \epsilon^d(\mathbf k) d^\dagger(\mathbf k) d(\mathbf k)+V_1 \sum_i c_{i;1}^\dagger d_i++V_2 \sum_i c_{i;2}^\dagger d_i+h.c.\notag\\
	&+\frac{U_1}{2}\sum_i n_{1;i}(n_{1;i}-1)+\frac{U_2}{2}\sum_i n_{2;i}(n_{2;i}-1)+U'\sum_i n_{1;i}n_{2;i}-2J_H \sum_i (\mathbf{S}_{1;i}\cdot \mathbf{S}_{2;i}+\frac{1}{4}n_{i;1}n_{i;2})
\end{align}
where $d$ is the hole operator for the orbital of $N_d$ element.  $c_1,c_2$ is the hole operators for the two eg orbitals of Ni. 

Unlike the model in Ref.~\onlinecite{ZXShen2019}, the above model  is not a simple Anderson model because of the inclusion of $V_2$ process. In the limit $J_H,U',U>>V_1,V_2$, a Kondo spin coupling between $N_i$ and $N_d$ spin can be derived from second order perturbation:
\begin{equation}
 	H_{kondo}=\tilde J \sum_i \vec S^{N_i}_i \cdot \vec S^{N_d}_i 
 \end{equation} 
where,
\begin{equation}
	\tilde J=\frac{2 V_1^2}{U_1}-2 V_2^2 \left(\frac{1}{U'-J_H}-\frac{1}{U'}\right)
\end{equation}

The super-exchange involving $V_2$ actually induces a ferromagnetic Kondo coupling because of the Hund's coupling. From DFT calculation, $V_1$ is very tiny \cite{ZXShen2019,2019arXiv190903942N}. Ref.\onlinecite{ZXShen2019} estimates $V_1 \approx 0.1 t_1$, which means $\tilde J \sim 0.01 J$. Hence the Kondo coupling between the Nd and Ni can be ignored.

We always have the constraint $n_{N_i}+n_{N_d}=1$.  Once there are electrons in $N_d$ orbital, $N_i$ site is self doped. In this case, we should expect a small doping $x$ even for undoped NdNiO$_2$ and the physics should be mainly governed by $t-J$ model in Eq.~\ref{eq:t_J_model_main} with small doping together with a small density of Nd electrons.  This self-doping effect is likely to be the reason why the parent compound is metallic and does not have magnetic order.  Besides, $n_{Nd}$ can in principle vary with temperature.  Decreasing of $n_{Nd}$ below $60$ K may be the reason why there is a change of Hall coefficient for $x=0.2$ and an upturn of resistivity for $x=0$.

In the limit $U, U-U', \epsilon_{dd}, J_H>>t$, we have a low energy model which extends our $t-J$ model:

\begin{equation}
	H=H_{t-J}+\sum_k \xi_{Nd}(\mathbf k)d^\dagger_\sigma(\mathbf k)d_\sigma(\mathbf k)+V \sum_i c^\dagger_i d_i+h.c.+J_{K} \vec S^s_i \cdot \vec S^{Nd}_i
\end{equation}
where $d^\dagger_i$ creates a hole for Nd orbital and $c^\dagger_i$ creates a spin-one doublon.  We need to include $J_K$ from $V_1$.

\end{document}